\documentstyle[aps,prl,multicol,psfig]{revtex}
\begin{document}
\input{epsf.tex}
\newcommand{\bea}{\begin{eqnarray}}
\newcommand{\eea}{\end{eqnarray}}
\newcommand{\beq}{\begin{equation}}
\newcommand{\enq}{\end{equation}}
\newcommand{\etal}{{\it et al.}}
\newcommand{\hPsi}{\hat{\Psi}}
\newcommand{\hn}{\hat{n}}
\newcommand{\rr}{{\bf r}}
\newcommand{\kk}{{\bf k}}
\newcommand{\xx}{{\bf x}}

\title{Surface-induced layer formation in polyelectrolytes}
\author{{ F. J. Solis and M. Olvera de la Cruz  }}
\address{ Department of Materials Science and Engineering, \\
	 Northwestern University, Evanston, Illinois 60208-3108. }

\maketitle

\begin{abstract} 
We analyze, 
by means of an RPA calculation, 
the conditions under which a mixture of oppositely charged 
polyelectrolytes can micro-segregate in the neighborhood of 
a charged surface creating a layered structure. 
A number of stable layers can be formed if the surface is 
sufficiently strongly charged even at temperatures 
at which the bulk of the mixture is homogeneous.
\end{abstract}

\begin{multicols}{2}

\section{Introduction}

It has been described in an article by Decher~\cite{sourc1,sourc2} a 
procedure for creating a coating for a (charged) surface that 
consist of alternating layers of two different, 
oppositely charged, 
polyelectrolyte homopolymers. 
The properties and uses of these systems has been explored by several 
groups ~\cite{emit8,mayes,optical}.
The basic technology for the creation of these coatings is easy to 
describe. The starting surface is charged (say, positively) 
and it is first dipped into a  solution of negatively charged 
homopolymers and then withdrawn from it. 
This forms a first layer since the polyelectrolyte is strongly attracted 
to the surface. 
The surface is washed and one gets rid of excess polymer 
chains  that are only loosely tied to the surface. 
The surface is now dipped in a solution of the second homopolymer 
that is positively charged. 
This second homopolymer is strongly attracted to the coated surface,  
which  is now effectively negatively charged. 
The process is repeated several times, and the layering is found to 
be stable. 
A more detailed exposition is presented in the original 
articles~\cite{sourc1,sourc2}. 
Figure 1 shows a schematic version of the layered system.

\begin{center}
\begin{minipage}[H]{3in}
\epsfxsize=3.0in \epsfbox{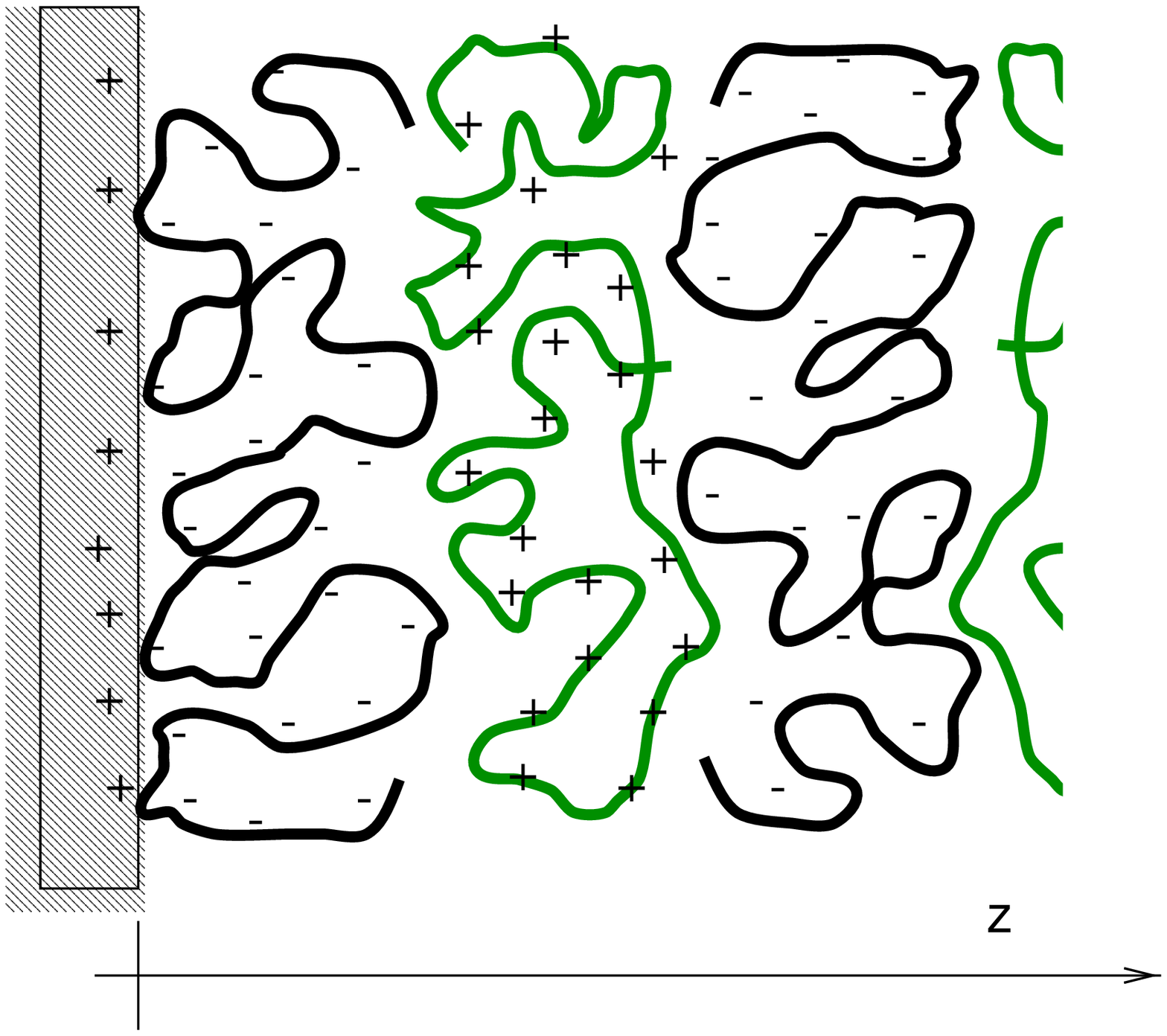}
\begin{figure}
\caption{ Scheme of the layered structure  
formed by a mixture of 
polyelectrolytes near a flat surface. The surface is charged positively, 
attracting a 
negatively charged polymer. The second layer consist of polymers 
positively charged, attracted to the first charged layer, and so forth. 
Counterions are not shown in the scheme.}
\end{figure}
\end{minipage}
\end{center}

It has been argued from the theoretical point of 
view,~\cite{monica,ruso1,ruso2,Doi}
that lamellar and other microphase structures can occur in mixtures
of oppositely charged polyelectrolytes. Usually, such  phases are associated
with block copolymers where chemical bonding is an impediment to 
macrophase separation. In charged polymer mixtures, the entropy
of mixing of the counterions for each of the homopolymer species provides
a huge barrier to macrophase formation, allowing instead for the possibility 
of microphases. In both cases, copolymers and polyelectrolytes, the 
separation is driven by the chemical repulsion between the 
species. 

	An interesting aspect of the above outlined procedure for manufacturing
the lamellar structure is that it would seem to work even at temperatures or 
for ranges of parameters at which, in principle, a phase 
separation should not be observed. In the following we will analyze the 
conditions under which we can 
expect to find such stable morphologies. Above the transition point,
an RPA approach will show that a finite  number of these layers should
be stable near the surface. 

	If the charged surface is immersed in a mixture of homopolymers,
it will drive the spontaneous creation of a few number of layers
that will then coexist with the bulk of the mixture. 
This  case is slightly different to the one resulting from the 
dip-and-wash mechanical layering procedure in which the surface is only 
in contact with solutions of homopolymers. Both systems, however,
should have similar stability conditions, the same relevant 
parameters, and generate comparable length scales for the layers. 
We will carry out our calculation for the case of the surface 
immersed in the polymer mixture, and consider the charged surface as a 
a large flat plate.

\section{RPA for polyelectrolyte mixtures}

	The fundamental fields for the description of the model 
system are the volume fraction of the homopolymers $\Phi_{A}$ and
$\Phi_{B}$. Both homopolymers will be considered to have the same monomer 
number $N$. A fraction $f$ of the monomers in each chain carry charges
$z_{A}$ and $z_{B}$ respectively and we will again consider only the 
most symmetric case $z=z_{A}=-{z_{B}}$. We consider the polymers to be 
dense with a negligible amount of solvent, and consider  the 
effective volume fraction occupied by the relative counterions very small. 
The local counterions  densities are $n_{A}$ and $n_{B}$. Counterions
for polymers $A$ and $B$ carry charges $-z_{A}$, and $-z_{B}$.

	The RPA approximation for a mixture of polyelectrolytes has 
been developed and studied by a number of 
groups~\cite{monica,ruso1,ruso2,Doi}. With a number of further 
simplifications (to be described later), a particularly useful form of the
RPA free energy functional has been written by Nyrkova {\etal}~\cite{Doi}. To 
their basic free energy, we add a term coupling the charge density to an 
external potential (that will be generated by the charged surface).
\bea
	F&=&\int d\rr 
	\Bigl\{ \frac{1}{N}\Phi_{A}\ln \Phi_{A} +
		\frac{1}{N}\Phi_{B}\ln \Phi_{B} + 
		\chi \Phi_{A}\Phi{B} 
	\nonumber \\
	&& + \frac{b}{36\Phi_{A}}(\nabla \Phi_{A})^{2}
		 + \frac{b}{36\Phi_{B}}(\nabla \Phi_{B})^{2}  
	\nonumber \\
	&& + n_{A}\ln n_{A} 
		 + n_{B}\ln n_{B} 
	\Bigr\}
	\nonumber \\
	&& 	+ \frac{Bj}{2} 
	     	\int d\rr\,d\rr'\,
		  \rho(\rr)\frac{1}{|\rr- \rr'|}\rho (\rr')
	   	+ \int d\rr\,eV(\rr)\rho(\rr) 
	\label{free0}
\eea
In this expression, all energies are measured in units of $k_{B}T$, and the
Kuhn length is $b$. The Bjerrum length is $Bj=e^{2}/(\epsilon k_{B}T)$. 
The interaction parameter $\chi$ is in 
this case positive, implying repulsion between species. The parameter
$\chi$ is considered close to, but smaller than,
the critical value required for the spontaneous creation of lamellar
(or another) microstructure. The prefactor for the gradient terms is such that 
if we turn off the  electrostatic 
interactions, the correlation functions will match the first terms of the 
Debye function for each type of chain, namely, in Fourier space 
$\left< \Phi_{\bf k} \Phi_{-\bf k} \right> = 
C(1+\frac{R^{2}}{3} k^{2}+\ldots)$, 
with the radius of gyration $R^{2}=b^2 N/6$. 
Lastly, the density of charged particles  $\rho$ is defined by
\beq
	\rho= z_{A}(f_{A}\Phi_{A}-n_{A})+z_{B}(f_{B}\Phi_{B}-z_{B}).
\enq 

	Thus, we have essentially a regular solution model in which Nyrkova
\etal~\cite{Doi} argue that 
there are no important corrections arising from fluctuations. Further, 
the intra-chain correlations are reduced to pure random walks, 
and only the first derivatives of the fields appear in the functional.
In reference \cite{monica} the full Debye function has been used to represent 
the intra-chain correlations, but the effect on the results here presented is
minor, as the corrections appear at wavelengths smaller than the 
resulting layer thickness.

	Since we are interested in the behavior of the mixture above
the demixing temperature, we assume a base state with homogeneous distributions
of homopolymers and counter-ions. The overall concentrations of polymers 
$\bar{\Phi}_{A}=\Phi_{0}$, $\bar{\Phi}_{B}=1-\Phi_{0}$ are considered as 
given, and the bulk concentrations of counter-ions 
are $\bar{n}_{i}=f\bar{\Phi}_{i}$, 
for each species. These overall concentrations should remain 
constant and the only important collective variables are the difference in 
the local concentrations between the species:
\bea
	\Psi &=& \Phi_{A} - \Phi_{B} -   (2\Phi_{0}-1) \\
	n    &=&    n_{A} -    n_{B} - f (2\Phi_{0}-1).
\eea
Since the volume fraction of the counterions is small compared to the volume
fraction of the polymers,  we set
\beq
	\Phi_{A}+\Phi_{B}=1,
\enq
to be satisfied everywhere.

	The effective RPA free energy in terms of the $\Psi,n$ variables is 
obtained by expanding all terms up to second order in the fields, and the 
result is 
\bea
	F&=&F_{0} + \frac{Bj}{2} \int d\rr \, d\rr' \,
	\frac{(f\Psi(\rr)-n(\rr))(f\Psi(\rr')-n(\rr'))}{|\rr-\rr'|} 
	 \nonumber \\
	&& +\int d\rr 
	 \frac{1}{8\Phi_{0}(1-\Phi_{0})N}
		\left( 	  \Psi^{2}
			+ \frac{b^2N}{18}(\nabla\Psi)^{2} 
			+ \frac{N}{f}n^{2}\right)
	\nonumber \\
	&&  -\int d\rr \, \frac{\chi}{4}\Psi^{2} 
		+ \int d\rr \, 	V(\rr)ez(f\Psi(\rr)-n(\rr)).
\eea
For our purposes, we need to Fourier transform all quantities 
($\hat{\Psi}(\kk)=\int d\rr \, e^{-i{\bf k \cdot  r}}\Psi(\rr)$, and so forth) 
and rewrite the free energy as:
\bea
	F&=&F_{0}+ 	\frac{Bj}{2} 
		\int \frac{d\kk}{(2\pi)^{3}}
			\frac{4\pi(f\hPsi-\hn)^{2}}{k^{2}} + 
	 \nonumber \\
	&& 
	\int \frac{d\kk}{(2\pi)^{3}}
		\frac{1}{8\Phi_{0}(1-\Phi_{0})N}
			\left(  \hPsi^{2} +
				\frac{k^{2}N}{18}\hPsi^{2} + 
				\frac{N}{f}\hn^{2} 	
			\right) 
		 \nonumber \\
	&& 
	-\int \frac{d\kk}{(2\pi)^{3}}\frac{\chi}{4}\hPsi^{2}  
	+\int d\kk\, 	ez\hat{V}(f\hPsi-\hn). 
	\label{ready}
\eea
	
\section{Response to external potentials and surface effects}

	For a given external potential $V(\rr)$, we can now calculate the
general response of the system. We shall work out the response when the
external potential originates from  a pair of parallel charged surfaces,
and then, in more detail, the case of only one bounding surface. 

\subsection*{A.General effect of an external potential}

	Minimization of the free energy, eq.\ref{ready}, with 
respect to the polymer and counterion densities, gives a pair
of coupled equations for $\hPsi$ and $\hn$. To present the solutions in a more
transparent manner, we will introduce a number of simplified and dimensionless
variables, namely:
\bea
	x&=&\frac{1}{2}r_{0}^{2}k^{2}=\frac{l^2 k^2}{18f}, \\
	s&=& \frac{1}{fN}\left[1-2\chi N\Phi(1-\Phi)\right], \\
	t&=& \frac{4 \pi \Phi (1-\Phi)}{9}\frac{Bj}{b}. \\
\eea
The variables $s$ and $t$ measure the strength of the interspecies and 
electrostatic interactions, while  the variable $x$ compares the wave-vectors
against the the screening length of the counter-ions $r_{0}=b(18f)^{-1/2}$.

	Now the solution to the system reads
\bea
	\hPsi(\kk) &=& R_{1}(\kk)\hat{V}(\kk)=\frac{
	\left[ 2 x  \Phi_{0} (1-\Phi_{0})   \right]
	(-e\hat{V(\kk)})}{D(\kk)} \\
	\hn(\kk) &=& R_{2}(\kk)\hat{V}(\kk)= \frac{
	\left[ 4 f \Phi_{0}(1-\Phi_{0}) x
	      (x+s)
	\right]
	e\hat{V(\kk)}}{D(\kk)} \label{solution}
\eea
where the denominator is 
\beq
	D(\kk) = x^2+ (t+s)x+ t(1+s),
\enq
and we have introduced a shorthand notation for the overall responses
$R_{1}$, and $R_{2}$.

\subsection*{B.Response of a mixture bounded by two surfaces}	

	We consider now a bulk of the mixture contained between two parallel
surfaces. One of the surfaces is charged positively, with a surface charge 
density $\sigma$.
 The more realistic condition for the second surface is to take it as 
uncharged. In fact, this second surface can be an imaginary surface coinciding 
with the free surface of the mixture (in contact with, say, air). The amount 
of charge in the second surface determines the electrostatic potential 
difference between the two surfaces, which  is the relevant driving force 
of the micro-segregation when we neglect non-linear terms. Thus, 
for simplicity, we consider instead the equivalent case in which 
one has two  oppositely charged surfaces with equal absolute value of 
charge density $\sigma/2$, separated by a distance $L$.
In the following, we take the $z$ axis to be normal to the 
surface, and set the origin in the left surface.

From elementary electromagnetism, the potential created by the charged  
surfaces is, for values of the $z$ coordinate  between $0$ and $L$ 
\beq
	V(\rr)= V(z)= \sigma L (\frac{L}{2}-z).
\enq
To use our Fourier transformed quantities, we extend the potential
to the full $z$ axis, by reflection at all planes with $z=nL$, with $n$ an 
integer. This allows us to obtain a suitable boundary behavior for the 
field $\Psi$, and reduces the problem 
to consider a Fourier Series for the potential 
in the region $(-L,L)$ (using the reflection condition). The expansion 
for the potential is $V=\sum{V_m e^{i m \pi z/L}}$, 
with coefficients
\beq
   	V_{m}=\cases{\frac{2 \sigma L}{\pi^2 m^2} & m odd \cr
			0 & m even. \cr } \label{cas}
\enq 
for 
The linear response to this potential is also expressed as a Fourier
Series, and the coefficients for the the composition $\Psi$ and the
counterion density $n$ are obtained immediately from the solution
in eq. \ref{solution} by evaluation at the wavenumbers 
$\kk_{m}, (k_{x}=m\pi/L, k_{y}=0, k_{z}=0)$:
\bea
	\Psi_{m} &=& R_{1}(\kk_{m})V_{m}, \label{series}\\ 
	\hn_{m} &=& R_{2}(\kk_{m})V_{m}. 
\eea

\subsection*{C. Response to the presence of one surface.}	
	
	The case of only one surface bounding the mixture is 
equivalent to the case considered in the previous subsection 
when we let the separation between the surfaces grow and effectively 
become infinite. The proper limit to consider for the 
Fourier transform  of the potential is
\beq
	\hat{V}(\kk)=2\sigma(2\pi)^2\delta(k_{x})\delta(k_{y})
	\frac{1}{k_{z}^2}. \label{infy}
\enq
It is convenient to introduce a new parameter proportional 
to the charge density,
\beq
	\gamma= 4e\sigma r_{o}^2 \Phi (1-\Phi),
\enq
which amounts to measure the surface charge density as the number
of charges in a square area of side $r_{o}$.

In the final result, all quantities depend only on the $z$ coordinate, 
and all relevant reference to 
the wave-vectors is only through  $x=k_{z}^2/18f$.
We obtain that the transform of the response in this limit 
can be expressed as 
\bea
	\hPsi(\kk)&=&(2\pi)^2\delta(k_{x})\delta(k_{y})
	\frac{-\gamma}{x^2+(s+t)x+t(1+s)}, \\
	\hn(\kk) &=& (2\pi)^2\delta(k_{x})\delta(k_{y})
	\frac{f(x+s)\gamma}{x^2+(s+t)x+t(1+s)}.
	 \label{solution2}
\eea
The denominator in these expressions is a real, even, quartic polynomial 
in $|\kk|$, and 
real quadratic in terms of $x$. 
	
	In this mean field approximation, there 
is a transition to a micro-segregated phase when the roots of the 
polynomial become real. This occurs when, for a given value of $t$, 
the $s$ variable becomes negative enough, namely, at  
\beq
	s_{r}=-2t^{1/2}+t,
\enq
and at that point, there are two pairs of double roots for the polynomial at
\beq
	x_{r}=\pm (t^{1/2}-t). \label{root}
\enq

	We are interested in the region above (temperature-wise), but 
close to the transition point. In such a region, the susceptibility to the 
influence of the surface is high, and yet, when the surface disappears there
will not be spontaneous micro-segregation. We can measure the distance 
from the transition point by the variable
\beq
	\epsilon= (s_{r}-s)^{1/2}.
\enq

In terms of this parameterization the response has a rather simple expression.
In real space we find that 
the field  $\Psi$  consist of decaying oscillations away from the $z=0$  plane,
({\it i.e.} the physical charged surface):
\beq
	\Psi(z)=-\frac{\gamma}{\epsilon x^{2}_{r}}
	\exp\left({-\frac{\epsilon}{r_{0}} z}\right)
	\cos\left(\frac{x_{r}^{1/2} z}{r_{0}}\right). 
	\label{mysolution}
\enq

\section{Layering}

	In the equilibrium compositional profile that we have 
obtained, eq.~\ref{mysolution}, a layer is represented by a peak (positive 
or negative) in the concentration variable. For a layer to be 
recognizable as such, the peak in the $\Psi$  field has to be comparable 
in magnitude to $1$ (in the case of symmetric mixtures) so that 
one of the components saturate a region in space thus forming a recognizable 
layer. Since the solution we have decays away from the surface, 
the number of layers that are formed is finite. It is clear that it
is of interest to maximize the number of layers that are stable 
near the surface. 
We shall consider three ways, suggested by our results, in 
which this effect can be achieved. 

	First, we note that that the field amplitude decays over a 
characteristic length $\Gamma=r_{0}/\epsilon$. If we tune the charge at the 
surface so that the strength of the electric field is just enough to saturate 
the composition at the surface so that $\Psi=1$ there, 
then layers will exist only up to a distance $\Gamma$ away from the surface. 
The thickness of the layers is given by the wavelength of the oscillations 
$\lambda=r_{0}/2 \pi x_{r}^{1/2}$, and therefore the number of layers formed
near the surface is 
\beq
	\frac{\Gamma}{\lambda}=\frac{2 \pi x_{r}^{1/2}}{\epsilon}.
\enq
To acquire the largest number possible of layers in this way, we would
like to make $\epsilon$ as small as possible. 
Thus one has to be as close as possible to the transition point for 
the number of layers to suitably increase.
  	An important issue arises here, in that while the layering system has 
been observed, the actual transition has not been observed, with the most 
simple explanation available being that the parameter region at which the 
system will undergo a transition is out of the reach of physically realizable
systems, but then, even the near-transition region ($\epsilon$ small), is
as well hard to reach. 

	A second idea for the creation of a good number of layers is 
over-saturation. Since  we are interested in attaining the $\Psi=1$ 
condition repeatedly, 
consider the case in which the source strength (the surface charge) is
so big that, mathematically, $\Psi$ will have to achieve values well beyond
$1$. To make sense of this solution, let us  neglect non-linear effects and 
simply assume that when the mathematical solution dictates  a value for 
$\Psi$ larger than $1$ the physical system simply saturates and takes
the  value of $1$ in that region.
Then, even if we go to distances well beyond $\Gamma$, the condition
$\Psi=1$ will be satisfied many times away from the surface. To obtain 
$m$ layers we need an amplitude $e^{m\lambda/\Gamma}$, so that
\beq
 m \sim \log (\gamma/\gamma_{0}), \label{layers}
\enq
where $\gamma_{0}$ is the effective surface charge density required to 
obtain saturation of one component at the surface, i.e. $\Psi=1$. 
 Schematically,
we have the case shown in Figure 2. 
While this is obviously a rather crude approximation, we should point out that,
at the linear response level, 
the thickness of the layers does not depend on the strength of the surface 
charge density and we can expect that the layers formed might still have
roughly such a characteristic size. A proper theoretical treatment of the 
problem should take into account the effects of strong segregation 
caused by the surface.

\begin{center}
\begin{minipage}[h]{3in}
\epsfxsize=3.2in \epsfbox{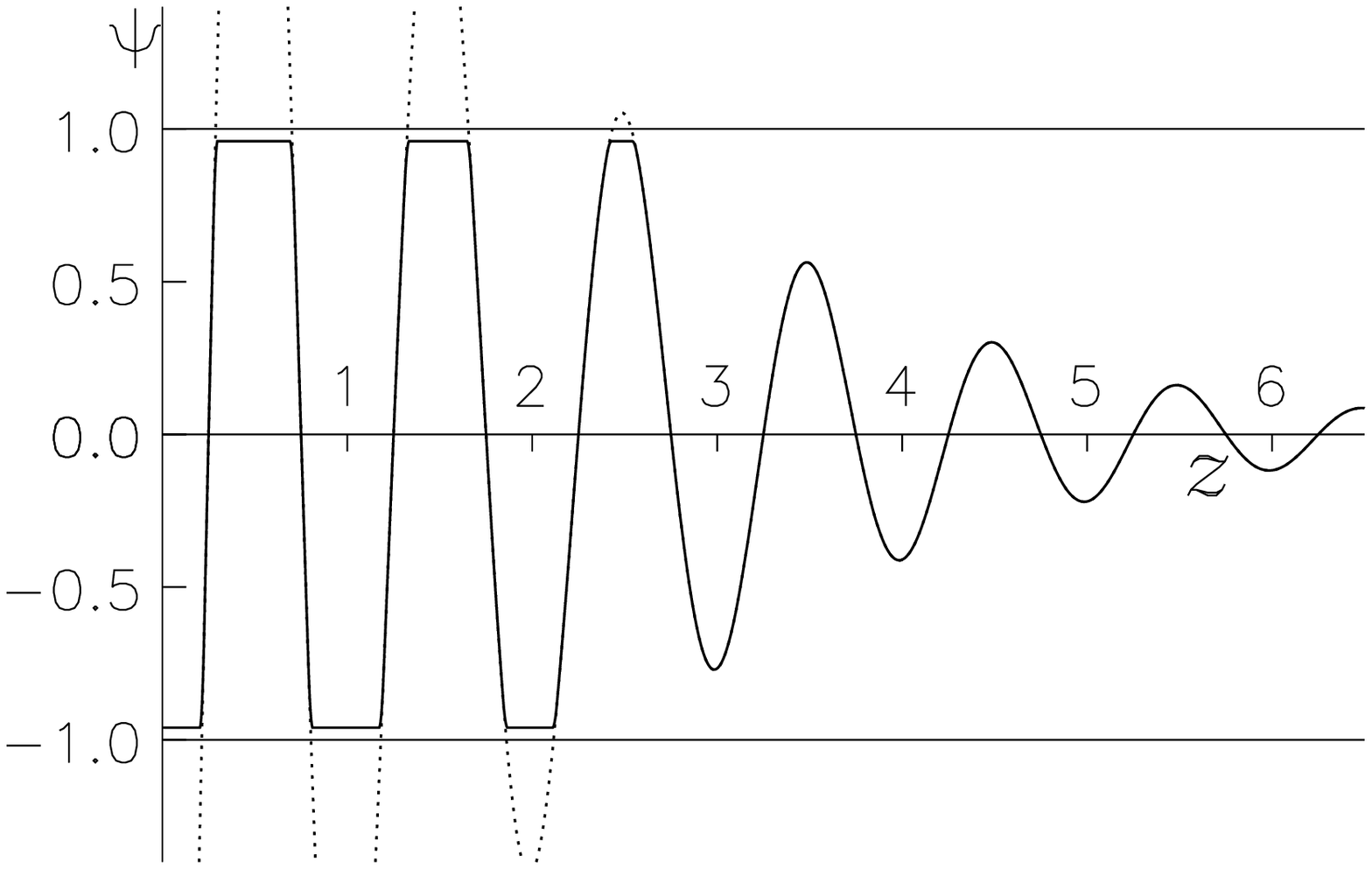}
\begin{figure}
\caption{
 The polymer composition profile $\Psi$, away from 
a charged surface. When the surface charge is large enough, the
amplitude as obtained in eq.\ref{mysolution} exceeds $1$. Constructing 
profile for the composition by a simple truncation of this solution
at the limiting values $\pm 1$, we obtain important compositional
fluctuations, even relatively far away from the surface.}
\end{figure}
\end{minipage}
\end{center}

	It is also important to note that with this kind of argument we 
could attempt to reconcile the experimental observation of the 
layering process, 
with the fact that at the same time there is not know trace of the 
non-forced microphase separation. 
This is, even when one operates far away from the critical point,
a sufficiently large driving force can show the traces 
of the micro-segregation. For simplicity, we have presented our 
results in a form valid only in the neighborhood of the transition.
For values of the parameters  moderately 
distant from  the transition point case, one should use the full
response, and look at the 
complex roots of the denominator $D$ to extract the wavelength, 
amplitude and decay length of the oscillations.

	Finally, we return to finite size effects. A third idea to
enhance the number of stable layers is to place the system in a finite 
geometry, as was described in section III. This is roughly equivalent to the 
experimental situation in which the layered system is taken away from the 
bulk of the mixture and has an overall finite thickness. 
	
	In the exploration of the 
limit when only one surface is present we have noticed that the 
oscillations of the homopolymers concentration decay relatively quickly.
This needs not to be the case when we have a finite distance between the
surfaces.
First, if the distance between them  is roughly about  twice the 
decay length $L \sim 2\Gamma$, we will have, in the non over-saturated case,
twice the number of layers expected from our estimate eq.~\ref{layers}, because
of the association of the layers with each of the two surfaces. Further, 
the (possibly) constructive interference of the oscillations will make 
the amplitudes larger around the mid-point between the surfaces, 
creating a stronger segregation pattern. Secondly, we can try to match a 
multiple of the preferred wavelength $m\lambda$ with the distance between 
surfaces. Clearly, this resonance maximizes the segregation of the species. 
In Figure 3 we show the response constructed from our 
result eq.~\ref{series}, for a  distance between surfaces equal to  $5.5$ 
times the wavelength $\lambda$ ($9$ layers).

\begin{center}
\begin{minipage}[h]{3in}
\epsfxsize=3.2in \epsfbox{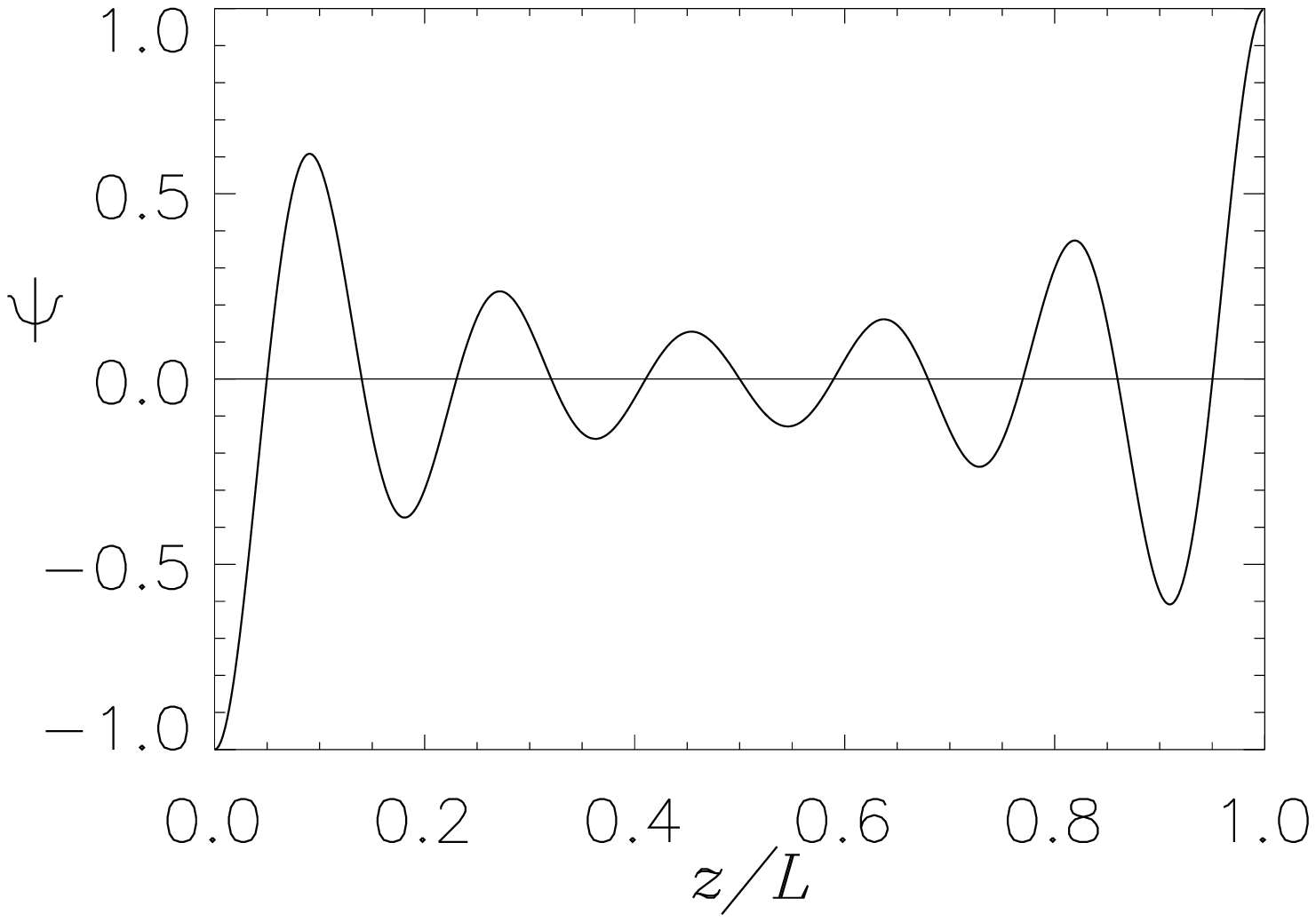}
\begin{figure}
\caption{
The polymer compositions profile 
$\Psi$ for a finite slab 
of polymer mixture of width $L=11\pi r_{0}/x^{1/2}_{r}=38r_{0}$. 
The values of the other basic parameters are $t=0.5$,  $s=-.91$.
}
\end{figure}
\end{minipage}
\end{center}

\section{Discussion}

	Our description of the response to the charged surface has 
been presented in terms of a set of reduced variables, $r_{0}, s, t$. 
But it is necessary
to comment on the values these quantities can achieve in actual systems. 
The simplest quantity is $t$. This is just a multiple of the Bjerrum length 
and we have it will always oscillate around a value of $1$, taking values of,
in some cases, $1/2$ for stiff polymers. 

	Next we must consider the values of $r_{0}$. This is perhaps the
more important consideration in the problem. We note that the root of the 
determinant $x$, in the vicinity of the critical point $x_{r}$, is 
roughly of the order of $t$, and from our 
previous paragraph, also of the order of $x_{r}$. Thus, in this region, 
$s$ will take values of order $1$. 
The oscillation frequency is then always of the same scale as the screening
length. Typically we will be interested in creating 
layers of thickness comparable, at the very least,  to the radius of 
gyration of a chain, $R^2 \sim N$, which together with 
$r_{0}^2 \sim 1/f $, leads to the condition
\beq
	fN \sim 1.
\enq
For this condition to be satisfied it is required that there should be  
only about $1$ free counterion per chain, or at least a very small number of 
them.  The reason while this condition might still be met in practice is the
possibility that many of the counterions are tightly bound to the chains in 
the melt, and $f$ represents only an effective number of free counterions.

	The reported findings of the Decher group~\cite{sourc1,sourc2},
seem consistent nevertheless with this picture. While they obtain 
mono-layers, the thickness of the layers is relatively small, about 
$100 A^{o}$, and the observed layering occurs also under
weakly charged conditions.

	In the neighborhood of the 
critical point of the mixture the parameter $s$ is also of order $-1$, which 
implies that 
\beq
	\chi N \sim fN \sim 1,
\enq
which simply implies that the segregation forces should be strong enough 
to compensate for the loss of translational entropy of the effectively 
free ions. This is not a stringent requirement and many known systems can 
satisfy the condition.

	Consider, finally, the charge density for the surface. 
Even when the systems is not too close to 
the microphase segregation point, and $\epsilon$ is not too small, 
the amplitude of the oscillation can be set to $1$ or be over-saturated
by increasing $\gamma$. If we
put $q$ charges in a surface area of size $R^2 \sim N$, we will have
$\gamma \sim q/fN \sim ~ q$. It is then clearly feasible to charge the 
surface with a large number of charges $q$ in the area occupied by one chain. 
The approach of obtaining a large number of layers using a very strong 
driving force seems then easily realizable. 

\section{Conclusions}

	We have explored the possibility of creating stable 
microstructures in polyelectrolytes by means of a strong 
external influence. While our qualitative results are consistent 
with experimental work, a more  precise comparison with experiments
cannot be done directly. We have used a number of effective parameters
some elucidation is required to relate them to the 
bare parameters of the system.  For example,it is necessary to develop
a relation between the original number of charges in every chain 
with the {\it effective} number of free counterions, and also to the 
effective value of the $\chi$ parameter, which should also be affected 
as the number of condensed counterions changes. These are subjects of 
forthcoming research.

	The only property that we have specified for the surface is 
its charge density, but is clear that it can be made of different materials,
and in some cases the surface can already have a previous polymeric coating. 
The layering scenario can then occur, for example, in the 
neighborhood of surfaces of charged polymer brushes ~\cite{russell}. 

	Another aspect of the problem that should be mention is the 
possibility of forcing 
different structures other than lamellar-like layers. When the 
volume fraction for one of the components is small, one
could expect that, as is the case in block copolymers, the 
 micro-segregation will create other structures, such
as spherical and cylindrical domains. Of course, there would also be an 
interplay between the geometry of the external charged surface and 
the preferred geometry of the segregated  system. 

\section*{Acknowledgments}
	
	The authors would like to thank Ann Mayes for useful conversations. 
	This work was sponsored by the National Science Foundation, 
grant DMR9807601.

\end{multicols}
\end{document}